\documentclass{cargese}\usepackage{graphicx}
\let\footnote\savefootnote
\let\footnotetext\savefootnotetext 
\let\footnoterule\savefootnoterule 
\normallatexbib
\newcommand{\beq}{\begin{equation}}
\newcommand{\eeq}{\end{equation}}
\newcommand{\beqs}{\begin{eqnarray}}
\newcommand{\eeqs}{\end{eqnarray}}
\newcommand{\dd}{{\mathrm d}}
\newcommand{\ee}{{\mathrm e}}

\begin{document}
\articletitle[Field Theory Duals from (Non)-Critical Type 0 Strings]
{Field Theory Duals from \\(Non)-Critical Type 0 Strings\footnote{Preprint
SISSA 110/99/EP.}}

\author{Dario Martelli\footnote{Talk presented at the NATO-ASI 
and TMR Summer School 'Progress in String Theory and M-Theory' Cargese, 
May 24 - June 5, 1999.}}

\affil{SISSA/ISAS, Via Beirut 2-4 34014 Trieste\\
and\\
INFN Sezione di Trieste, Italy}        
\email{dmartell@sissa.it}

\begin{abstract}
We review some aspects of Polyakov's proposal for constructing 
nonsupersymmetric field theories from non-critical Type 0 string theory.  
\end{abstract}

Getting some insight into nonsupersymmetric field theory via AdS/CFT
approach is a difficult task to achieve. 
There are different proposals to address 
this issue, dealing with possible mechanisms to break supersymmetry.
Polyakov in \cite{Polyakov:a} has proposed considering a nonsupersymmetric
string theory with non-chiral GSO projection, known as Type 0   
\cite{Dixon,Seiberg}. This proposal was elaborated further in a series of
papers starting from \cite{Klebanov:a}. In this short note we present
the results obtained in \cite{Ferretti:b,Ferretti:a}.  

Type 0 string theories are purely bosonic strings with modular invariant
partition function and four sectors whose low laying fields 
are here summarized\footnote{The two entries in parenthesis refer
respectively to left and right movers and the signs correspond to the
choice of GSO projection. The upper (lower) signs in the RR sectors
define Type 0A (0B) theory.}
\begin{center}
\begin{tabular}{c|cccc}
 & $(NS+,NS+)$ & $(NS-,NS-)$ & $(R+,R\mp)$ & $(R-,R\pm)$ \\
\hline 
{\bf 0A} & $\Phi ,B_{\mu\nu} ,g_{\mu\nu}$ & $T$ &
$A^{(1)}_{\mu} A^{(1)}_{\mu\nu\rho}$ &
$A^{(2)}_{\mu} A^{(2)}_{\mu\nu\rho}$ \\
{\bf 0B} & $\Phi ,B_{\mu\nu} ,g_{\mu\nu}$ & $T$ & 
$A^{(1)} A^{(1)}_{\mu\nu} A^{(+)}_{\mu\nu\rho\sigma}$ & 
$A^{(2)} A^{(2)}_{\mu\nu} A^{(-)}_{\mu\nu\rho\sigma}$ \\
\end{tabular}
\end{center}
There is a corresponding doubled set of D-branes coupling to RR-fields.

As pointed out in \cite{Polyakov:a} there are no open string tachyons
on the world-volume of these branes, while there is a perturbative closed
string tachyon that renders the Minkowski vacuum unstable. 
Nevertheless one should 
regard this as an indication that ten dimensional flat background is not 
stable, while there should exist other vacua in which the theory makes sense.
AdS space seems in this respect a good candidate in that it allows for 
tachyonic modes.

${\cal N}=(1,1)$ supersymmetry on the world-sheet makes these theories similar
in some 
respect to supersymmetric Type II. For example all tree level correlators 
of vertex 
operators of (NS+,NS+) and (R+,R+) fields are the same as Type II. 
Using these and other
properties it is possible to derive an expression for the effective gravity
action \cite{Klebanov:a} that, split in the NSNS and RR contributions,
reads:
\beqs
S_{\rm NSNS} = \int \dd^{10}x\;\sqrt{-g}\left\{\ee^{-2\Phi}\left(
R -\frac{1}{12} |\dd B|^2 + 4 |\dd\Phi|^2 - 
\frac{1}{2}|\dd T|^2 - V(T) \right) \right\} \nonumber
\eeqs
\beqs
S_{\rm RR} = \int \dd^{10}x\;\sqrt{-g}\left\{f(T)|F_{p+2}|^2+\cdots+
({\rm CS\ terms})\right\}~.\nonumber
\eeqs
The main novelties are coming from the tachyon couplings. In particular, 
there is
a potential $V(T)$ which is an even function of the tachyon field, 
as well as functions $f(T)$ multiplying RR terms, that can be
worked out perturbatively.

The original proposal in \cite{Polyakov:a} was to consider a string theory in 
dimension $d<10$. Even if a microscopic description of string theory out of 
criticality is still far, there are at least indications for a possible 
extension of Type 0 theories in non-critical dimensions:
\begin{itemize}
\item
a diagonal partition function, whose modular invariance doesn't rely 
on $d=10$ (as opposed to Type II theories)
\item
the tachyon should condense, providing an effective central charge 
$c_{\rm eff}\sim V(\langle T\rangle)$. It doesn't seem unnatural to shift  
$c_{\rm eff}$ by the central charge deficit $(10-d)$ 
\end{itemize}
$c_{\rm eff}$ provides a tree level cosmological constant in the low energy
theory,
and this agrees with the expectation that the inconsistencies possibly arise
only in flat background.

We work at the level of the effective gravity action. In $d<10$ one
should guess the field content of the theory and write down the relative
action. In \cite{Ferretti:b} we assume the NSNS sector (gravity + tachyon) 
is always 
present and the RR sector is worked out on group theory grounds, considering
tensor products of $SO(d-2)$ spinors. In five dimensions for instance, 
${\bf 2} \times {\bf 2}={\bf 1}+{\bf 3}$, and one is led to include
a scalar potential $A$ and a vector $A_{\mu}$. We further assume the existence
of a $4$-form potential to accommodate the would be D3-brane: This really 
amounts to consider massive gravity.    

With these assumptions, the equations of motion following from the
relative action
have interesting solutions, whose interpretation may give some insight
into their field theory duals, and eventually provide hints
in favor of the consistency of either critical or even non-critical
Type 0 string theory.
 
The relevant piece of the $d$-dimensional action, for the ansatz that 
we consider is
\footnote{Uppercase indices run from $1$ to $d$, while Greek and Latin 
indices run from $1$ to $p+2$ and from $p+3$ to $d$ respectively.}
\beqs
S &=& \int \dd^dx \sqrt{-g}~ \Bigg\{ R - \frac{1}{2} (\partial_{M}\Phi)^2 
- \frac{1}{2} (\partial_{M}T)^2 - V(T)~\ee^{\sqrt{\frac{2}{d-2}}\Phi} \nonumber \\ & & \qquad
- \frac{1}{2~(p+2)!}~ f(T)~\ee^{\frac{1}{2}\sqrt{\frac{2}{d-2}}(d-2p-4)\Phi}~ \Big(F_{M_1 \cdots M_{p+2}}\Big)^2  
\Bigg\},\nonumber
\eeqs
where $V(T)=-10+d-\frac{d-2}{8}T^2+\cdots$ is the tachyon potential, including 
central charge deficit.

Nonzero fields are the metric, constant dilaton ($\Phi_0$) and tachyon 
($T_0$), and a RR ($p+2$)-form field strength: 
\beqs
     R_{\mu\nu\rho\lambda}=-\frac{1}{R^2_0}\left(g_{\mu\rho}g_{\nu\lambda}
         -g_{\mu\lambda}g_{\nu\rho} \right)~,\quad
     R_{ijkl}=+\frac{1}{L^2_0}\left(g_{ik}g_{jl}-g_{il}g_{jk} 
         \right)~, \nonumber
\eeqs
\beqs
F_{\mu_1 \cdots \mu_{p+2}} &=& F_0~ \sqrt{-g_{(p+2)}}~ 
\epsilon_{\mu_1 \cdots \mu_{p+2}}~.\nonumber 
\eeqs
With such an ansatz the equations of motion become a set of algebraic
equations. The tachyon VEV is determined implicitly by the following equation
\beqs
\frac{f'(T_0)}{f(T_0)} &=& \frac{1}{2}(d-2p-4)~ \frac{V'(T_0)}{V(T_0)}~.
\nonumber 
\label{tnaught}
\eeqs
Now, without a precise knowledge of the functions $f(T)$ and $V(T)$ one
cannot infer whether or not it admits solutions. 
One should really assume it has, and extract some information. 

The remaining equations fix the value
of the radii of the two maximally symmetric spaces and that of the dilaton.
It turns out that such solution is a 
${\mathrm AdS}_{p+2} \times {\mathrm S}^{d-p-2}$ space, with tachyon
VEV, and fixed 't Hooft coupling $\lambda=e^{\Phi_0}N$.\footnote
{N is the number of branes, which has to be evaluated in the string frame.}

Notice that string loop corrections are suppressed in the large N limit, while
$\alpha'$-corrections are important because the curvature is $O(1)$.

By the AdS/CFT correspondence the field theory dual of this solution
should be at a conformal point. However it is difficult to make contact
with perturbative field theory because $\lambda\sim O(1)$. Nevertheless
one can still get additional information applying the correspondence.
It is in fact possible to count the number of degrees of freedom  
that should live in the theory dual to this background \cite{Ferretti:a}.

Consider a thermal deformation of the solution. Identifying the Hawking  
temperature with the finite temperature of the field theory one can
compute the entropy. By either computing the free energy from the Euclidean
action, or computing the area of the horizon,
it turns out that it has the following behavior
\beqs
S\sim N^2V_pT_H^p \nonumber
\eeqs
for any value of $p$. This is an indication that the dual field theory should
have $N^2$ degrees of freedom, and could be YM theory in some non Gaussian 
limit. Notice
the different scaling power in the analogous relation one gets from evaluating
the entropies of black M2 and M5 branes ($3/2$ and 3 respectively).

We conclude pointing out that this kind of approach is potentially 
predictive. Consider in fact the formula for scaling dimensions of
dual operators\footnote{These results are obtained in the case d=p+2.}
\beqs
\Delta &=& \frac{(d-1)+\sqrt{(d-1)+4m^2R_0^2}}{2}\nonumber\\
m^2R_0^2 &=& d(d-1)\, \Bigg(1 + \frac{\tau}{2} 
\pm \frac{1}{2}\sqrt{\tau^2+(2d-4)\frac{V'(T_0)^2}{V(T_0)^2}}\Bigg)\nonumber
\eeqs
\beqs
\qquad {\rm with} \qquad 
\tau = d\frac{V'(T_0)^2}{V(T_0)^2} -\frac{2}{d}\frac{f''(T_0)}{f(T_0)}
-\frac{V''(T_0)}{V(T_0)}-1~.\nonumber
\eeqs

First, note that the tachyon VEV $T_0$ behaves as a ``bare'' quantity: 
it does not
enter in determining physical quantities. It is very much as in the
renormalization group. One can do a field redefinition, this will shift
$T_0$, without affecting $\lambda$, $R_0$, $\Delta$.

Then, the scaling dimensions depend on a {\it finite} number of 
parameters and with a good guess
on the field theory side, one in principle should be able to predict an 
infinite tower of dimensions from KK analysis. Moreover, some control on the
theory at this conformal point may shed light on Type 0 string theory and/or on
non-critical string theory via AdS/CFT correspondence.

Further details concerning this short note can be found in 
\cite{Ferretti:b,Ferretti:a}.      

\begin{acknowledgments}
I wish to thank G. Ferretti for reading the manuscript and J. Kalkkinen 
for useful discussions. 
I also thank the organizers of the school for providing 
a nice environment in which to learn and discuss. 
This work was supported in part by European Union TMR program CT960045.
\end{acknowledgments}

\end{document}